\documentclass[10pt,twocolumn,letterpaper]{article}

\usepackage{titling}
\usepackage{listings}

\usepackage{cvpr}
\usepackage{times}
\usepackage{multicol}

\usepackage{epsfig}
\usepackage{graphicx}
\usepackage{amsmath}
\usepackage{mathrsfs}
\usepackage{amssymb}
\usepackage{subfigure}
\usepackage{enumitem}

\usepackage{algorithm}
\usepackage{algorithmic}

\usepackage[title]{appendix}

\usepackage{xcolor}

\usepackage{macros}

\newcommand{\code}[1]{\texttt{#1}}

\usepackage[pagebackref=true,breaklinks=true,letterpaper=true,colorlinks,bookmarks=false]{hyperref} 

\usepackage{footmisc}

\cvprfinalcopy 

\def\cvprPaperID{****} 

\begin{document}

\begin{multicols}{1} 
\title{Learned Video Compression}

\author{Oren Rippel,\ \ Sanjay Nair,\ \ Carissa Lew,\ \ Steve Branson,\ \ Alexander G. Anderson,\ \ Lubomir Bourdev\\
WaveOne, Inc.\\
{\tt\small \{oren, sanjay, carissa, steve, alex, lubomir\}@wave.one}
}

\maketitle
\end{multicols}


\begin{abstract}
We present a new algorithm for video coding, learned end-to-end for the low-latency mode. In this setting, our approach outperforms all existing video codecs across nearly the entire bitrate range. To our knowledge, this is the first ML-based method to do so.

We evaluate our approach on standard video compression test sets of varying resolutions, and benchmark against all mainstream commercial codecs, in the low-latency mode. On standard-definition videos, relative to our algorithm, HEVC/H.265, AVC/H.264 and VP9 typically produce codes up to 60\% larger. On high-definition 1080p videos, H.265 and VP9 typically produce codes up to 20\% larger, and H.264 up to 35\% larger. Furthermore, our approach does not suffer from blocking artifacts and pixelation, and thus produces videos that are more visually pleasing.

We propose two main contributions. The first is a novel architecture for video compression, which (1) generalizes motion estimation to perform any learned compensation beyond simple translations, (2) rather than strictly relying on previously transmitted reference frames, maintains a state of arbitrary information learned by the model, and (3) enables jointly compressing all transmitted signals (such as optical flow and residual).

Secondly, we present a framework for ML-based spatial rate control --- a mechanism for assigning variable bitrates across space for each frame. This is a critical component for video coding, which to our knowledge had not been developed within a machine learning setting.
\end{abstract}


\section{Introduction}
Video content consumed more than 70\% of all internet traffic in 2016, and is expected to grow threefold by 2021 \cite{CISCOForecast}. At the same time, the fundamentals of existing video compression algorithms have not changed considerably over the last 20 years \cite[\dots]{vanne2012comparative,pourazad2012hevc,ohm2012comparison}. While they have been very well engineered and thoroughly tuned, they are hard-coded, and as such cannot adapt to the growing demand and increasingly versatile spectrum of video use cases such as social media sharing, object detection, VR streaming, and so on.

Meanwhile, approaches based on deep learning have revolutionized many industries and research disciplines. In particular, in the last two years, the field of image compression has made large leaps: ML-based image compression approaches have been surpassing the commercial codecs by significant margins, and are still far from saturating to their full potential (survey in Section \ref{sec:related_work}).

The renaissance of deep learning has further catalyzed the proliferation of architectures for neural network acceleration across a spectrum of devices and machines. This hardware revolution has been increasingly improving the performance of deployed ML-based technologies --- rendering video compression a prime candidate for disruption.

In this paper, we introduce a new algorithm for video coding. Our approach is learned end-to-end for the low-latency mode, where each frame can only rely on information from the past. This is an important setting for live transmission, and constitutes a self-contained research problem and a stepping-stone towards coding in its full generality. In this setting, our approach outperforms all existing video codecs across nearly the entire bitrate range.

\begin{figure*}[t!]
\setlength{\columnsep}{0pt}
\centering
\begin{minipage}{\textwidth}
\vspace{-0.3in}
  \centering
  \raisebox{-0.4\height}{\includegraphics[width=1.5in,trim=0cm 0cm 0cm 0cm,clip]{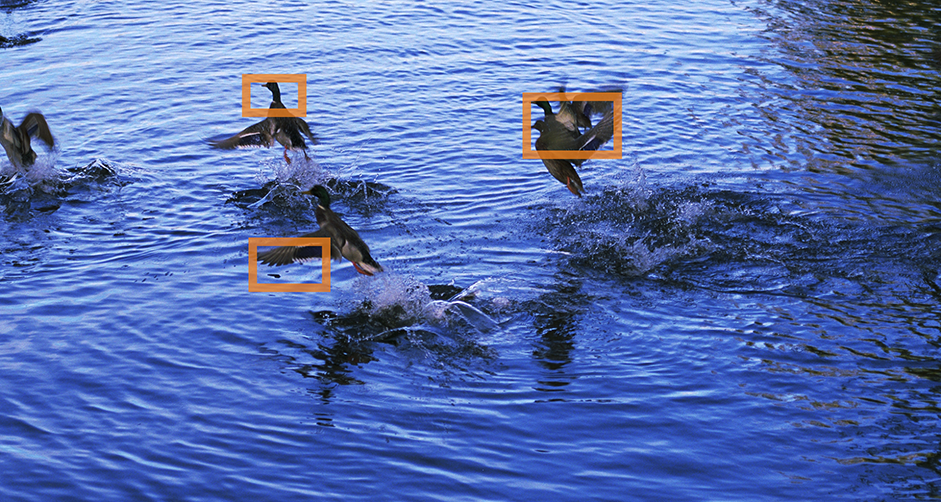}}
  \raisebox{-0.5\height}{\includegraphics[height=1.25in,trim=0cm 0cm 0cm 0cm,clip]{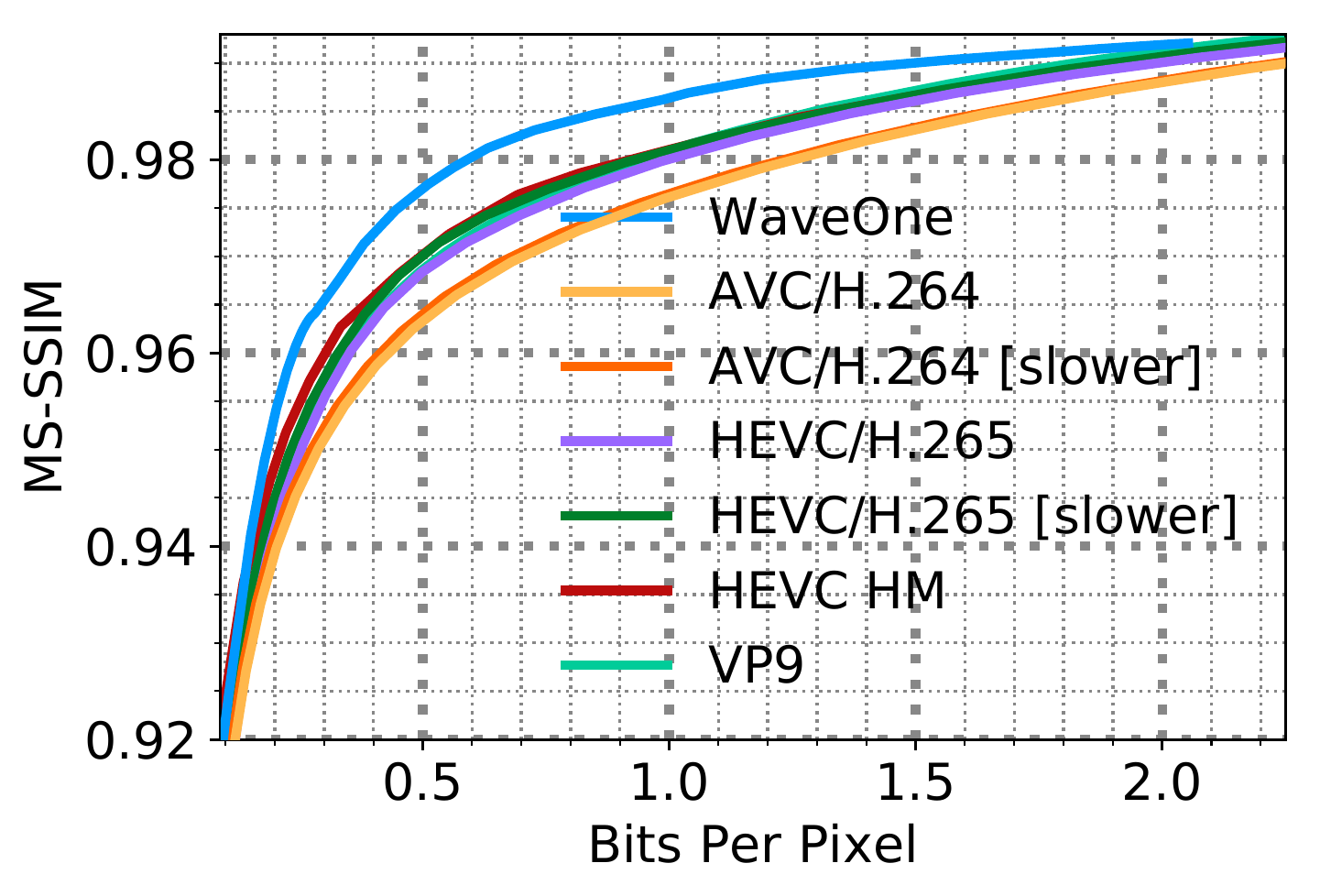}}
  \raisebox{-0.4\height}{\includegraphics[width=1.5in,trim=0cm 0cm 0cm 0cm,clip]{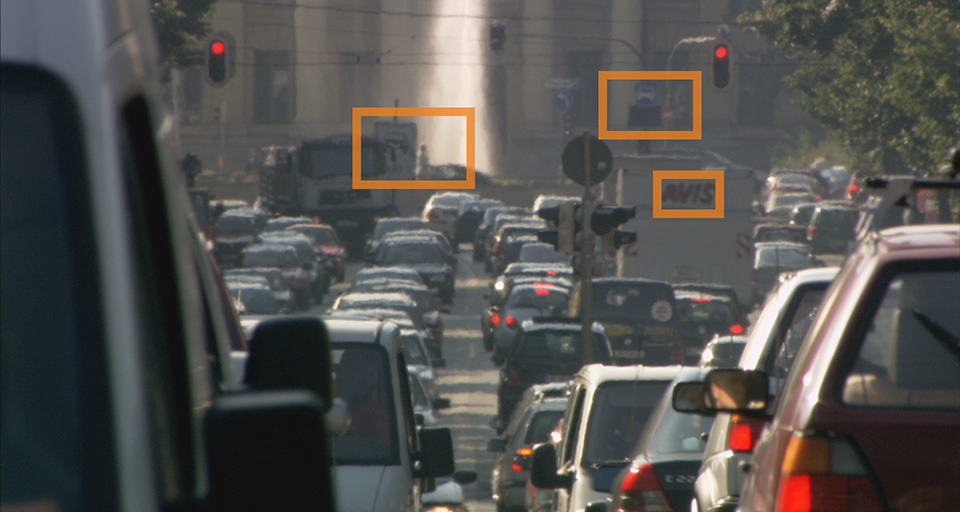}}
  \raisebox{-0.5\height}{\includegraphics[height=1.25in,trim=0cm 0cm 0cm 0cm,clip]{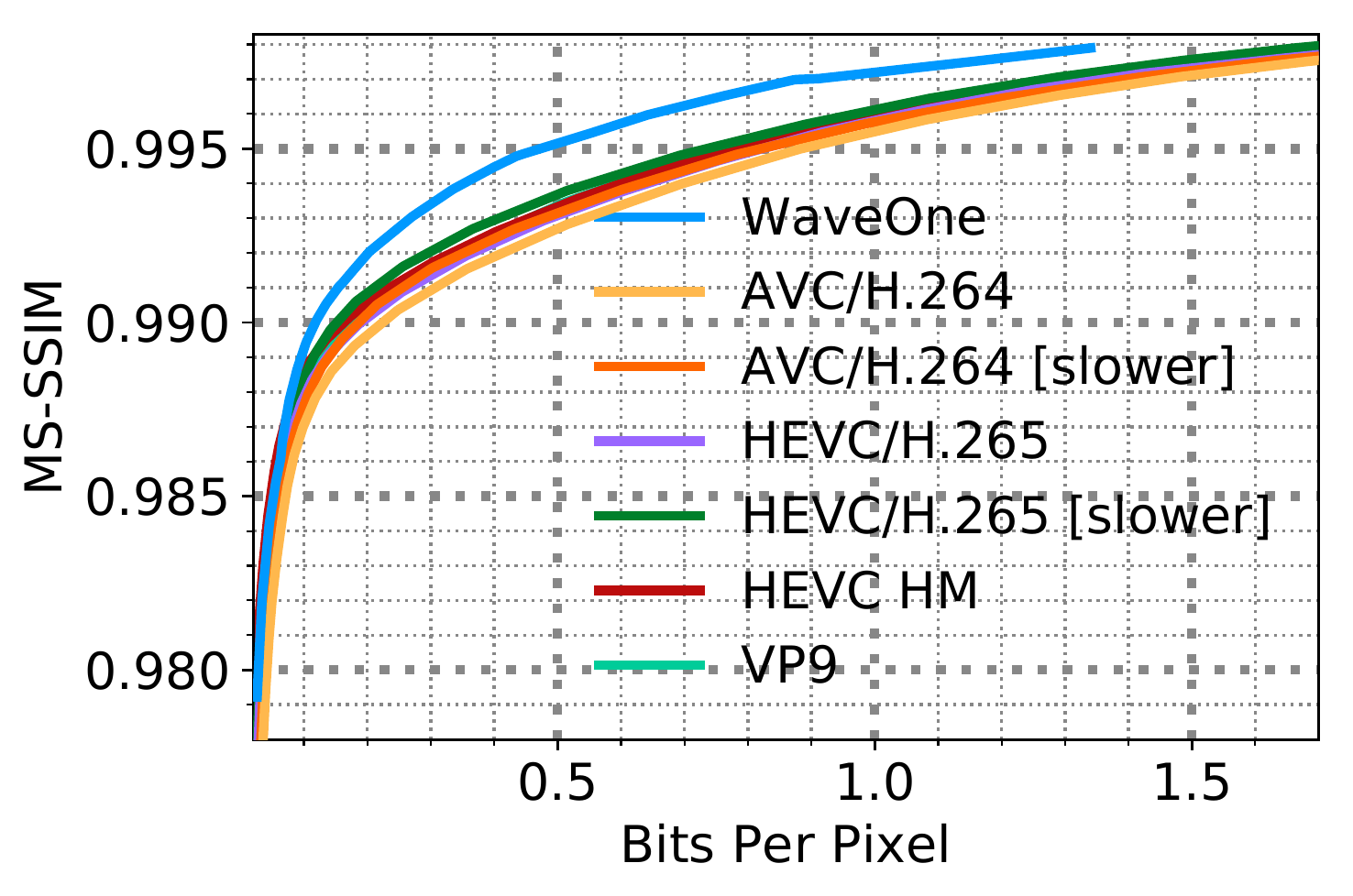}}
\end{minipage}
\vspace{-0.12in}
\begin{multicols}{6}
\centering
\includegraphics[width=0.16\textwidth,trim=0cm 0cm 0cm 0cm,clip]{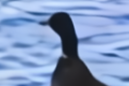}
\includegraphics[width=0.16\textwidth,trim=0cm 0cm 0cm 0cm,clip]{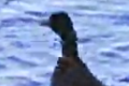}
\includegraphics[width=0.16\textwidth,trim=0cm 0cm 0cm 0cm,clip]{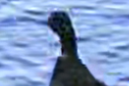}
\includegraphics[width=0.16\textwidth,trim=0cm 0cm 0cm 0cm,clip]{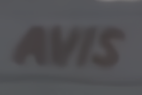}
\includegraphics[width=0.16\textwidth,trim=0cm 0cm 0cm 0cm,clip]{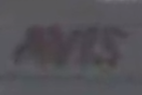}
\includegraphics[width=0.16\textwidth,trim=0cm 0cm 0cm 0cm,clip]{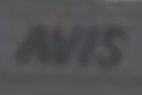}
\end{multicols}
\vspace{-0.32in}
\begin{multicols}{6}
\centering
\includegraphics[width=0.16\textwidth,trim=0cm 0cm 0cm 0cm,clip]{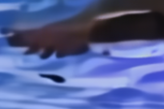}
\includegraphics[width=0.16\textwidth,trim=0cm 0cm 0cm 0cm,clip]{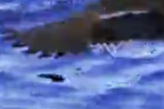}
\includegraphics[width=0.16\textwidth,trim=0cm 0cm 0cm 0cm,clip]{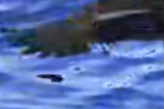}
\includegraphics[width=0.16\textwidth,trim=0cm 0cm 0cm 0cm,clip]{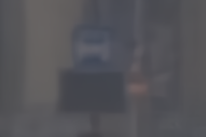}
\includegraphics[width=0.16\textwidth,trim=0cm 0cm 0cm 0cm,clip]{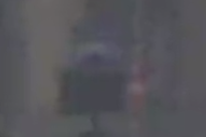}
\includegraphics[width=0.16\textwidth,trim=0cm 0cm 0cm 0cm,clip]{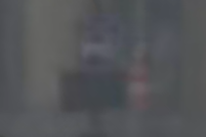}
\end{multicols}
\vspace{-0.32in}
\begin{multicols}{6}
\centering
\includegraphics[width=0.16\textwidth,trim=0cm 0cm 0cm 0cm,clip]{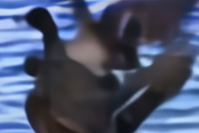}\\
{\bf Ours}\\
0.0570 BPP
\includegraphics[width=0.16\textwidth,trim=0cm 0cm 0cm 0cm,clip]{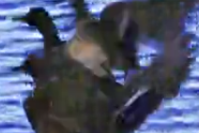}\\
{\bf AVC/H.264}\\
0.0597 BPP
\includegraphics[width=0.16\textwidth,trim=0cm 0cm 0cm 0cm,clip]{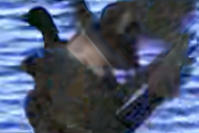}\\
{\bf AVC/H.265}\\
0.0651 BPP
\includegraphics[width=0.16\textwidth,trim=0cm 0cm 0cm 0cm,clip]{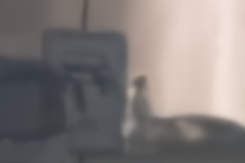}\\
{\bf Ours}\\
0.0227 BPP
\includegraphics[width=0.16\textwidth,trim=0cm 0cm 0cm 0cm,clip]{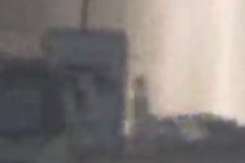}\\
{\bf AVC/H.264}\\
0.0247 BPP
\includegraphics[width=0.16\textwidth,trim=0cm 0cm 0cm 0cm,clip]{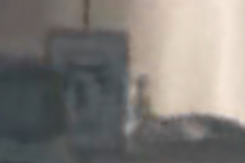}\\
{\bf AVC/H.265}\\
0.0231 BPP
\end{multicols}
\caption{Examples of reconstructions by different codecs for the same bits per pixel (BPP) value. Videos taken from the Xiph HD library\protect\footref{xiph}, commonly used for compression evaluation. Comprehensive benchmarking results can be found in Section \ref{sec:results}. {\bf Top left:} raw input frame, with boxes around areas zoomed-in on. {\bf Top right:} rate-distortion curves for each video. {\bf Bottom rows:} crops from the reconstruction by each codec for visual comparisons of fine details (better viewed electronically).}
\setlength{\columnsep}{22.58437pt}
\label{fig:intro_examples}
\end{figure*}

We thoroughly evaluate our approach on standard datasets of varying resolutions, and benchmark against all modern commercial codecs in this mode. On standard-definition (SD) videos, relative to our algorithm, HEVC/H.265, AVC/H.264 and VP9 typically produce codes up to 60\% larger. On high-definition (HD) 1080p videos, H.265 and VP9 typically produce codes up to 20\% larger, and H.264 up to 35\% larger. Furthermore, our approach does not suffer from blocking artifacts and pixelation, and thus produces videos that are more visually pleasing (see Figure~\ref{fig:intro_examples}).

In Section \ref{sec:existing_codecs}, we provide a brief introduction to video coding in general. In Section \ref{sec:contributions}, we proceed to describe our contributions. In Section \ref{sec:related_work} we discuss related work, and in Section \ref{sec:paper_organization} we provide an outline of this paper.

\subsection{Video coding in a nutshell}\label{sec:existing_codecs}
\subsubsection{Video frame types}
Video codecs are designed for high compression efficiency, and achieve this by exploiting spatial and temporal redundancies within and across video frames (\cite{wiegand2003overview,vetro2003video,pourazad2012hevc,mukherjee2013latest} provide great overviews of commercial video coding techniques). Existing video codecs feature 3 types of frames: 
\begin{enumerate}
\item I-frames ("intra-coded"), compressed using an image codec and do not depend on any other frames;
\item P-frames ("predicted"), extrapolated from frames in the past; and
\item B-frames ("bi-directional"), interpolated from previously transmitted frames in both the past and future.
\end{enumerate}
While introducing B-frames enables higher coding efficiency, it increases the latency: to decode a given frame, future frames have to first be transmitted and decoded.

\vspace{-0.05in}
\subsubsection{Compression procedure}
\vspace{-0.05in}
In all modern video codecs, P-frame coding is invariably accomplished via two separate steps: (1) motion compensation, followed by (2) residual compression.
\vspace{-0.05in}

\vspace{-0.05in}
\paragraph{Motion compensation.} The goal of this step is to leverage temporal redundancy in the form of translations. This is done via block-matching (overview at \cite{manikandan2014new}), which reconstructs the current target, say $\rmbx_t$ for time step $t$, from a handful of previously transmitted \emph{reference frames}. Specifically, different blocks in the target are compared to ones within the reference frames, across a range of possible displacements. These displacements can be represented as an optical flow map $\rmbf_t$, and block-matching can be written as a special case of the \emph{flow estimation} problem (see Section \ref{sec:flow_estimation}). In order to minimize the bandwidth required to transmit the flow $\rmbf_t$ and reduce the complexity of the search, the flows are applied uniformly over large spatial blocks, and discretized to precision of half/quarter/eighth-pel.
\vspace{-0.05in}

\vspace{-0.05in}
\paragraph{Residual compression.} Following motion compensation, the leftover difference between the target and its motion-compensated approximation $\rmbm_t$ is then compressed. This difference $\bDelta_t = \rmbx_t - \rmbm_t$ is known as the \emph{residual}, and is independently encoded with an image compression algorithm adapted to the sparsity of the residual. 

\subsection{Contributions}\label{sec:contributions}
This paper presents several novel contributions to video codec design, and to ML modeling of compression:

\vspace{-0.05in}
\paragraph{Compensation beyond translation.} Traditional codecs are constrained to predicting temporal patterns strictly in the form of motion. However, there exists significant redundancy that cannot be captured via simple translations. Consider, for example, an out-of-plane rotation such as a person turning their head sideways. Traditional codecs will not be able to predict a profile face from a frontal view. In contrast, our system is able to learn arbitrary spatio-temporal patterns, and thus propose more accurate predictions, leading to bitrate savings.

\vspace{-0.05in}
\paragraph{Propagation of a learned state.} In traditional codecs all ``prior knowledge'' propagated from frame to frame is expressed strictly via reference frames and optical flow maps, both embedded in raw pixel space. These representations are very limited in the class of signals they may characterize, and moreover cannot capture long-term memory. In contrast, we propagate an arbitrary \emph{state} autonomously learned by the model to maximize information retention.

\vspace{-0.05in}
\paragraph{Joint compression of motion and residual.} Each codec must fundamentally decide how to distribute bandwidth among motion and residual. The optimal tradeoff between them is different for each frame. In traditional methods, the motion and residual are compressed separately, and there is no easy way to trade them off. Instead, we jointly compress the compensation and residual signals, using the same bottleneck. This further eliminates redundancies among them, and allows our network to learn how to distribute the bitrate among them as function of frame complexity.

\begin{figure}[t]
\centering
\subfigure[\small H.265 motion vectors.]{
  \includegraphics[width=0.48\columnwidth,trim=0cm 0cm 0cm 0cm,clip]{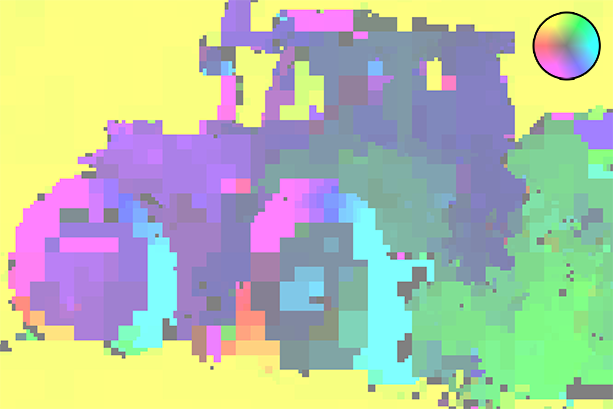}
}\hspace{-0.04in}
\subfigure[\small Our optical flow.]{
  \includegraphics[width=0.48\columnwidth,trim=0cm 0cm 0cm 0cm,clip]{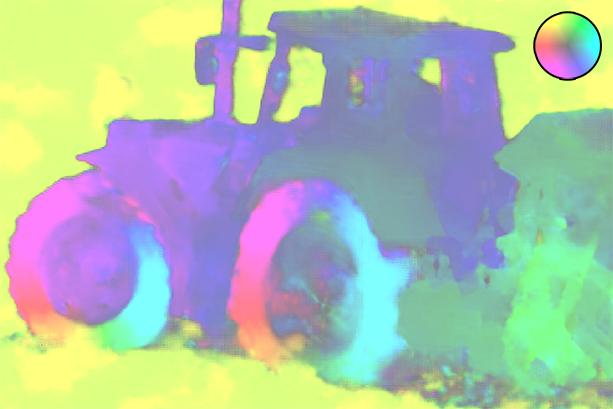}
}

\caption{Optical flow maps for H.265 and our approach, for the same bitrate. Traditional codecs use a block structure to represent motion, and heavily quantize the motion vectors. Our algorithm has the flexibility to represent arbitrarily sophisticated motion.}
\label{fig:flows_comparison}
\end{figure}

\vspace{-0.05in}
\paragraph{Flexible motion field representation.} In traditional codecs, optical flow is represented with a hierarchical block structure where all pixels within a block share the same motion. Moreover, the motion vectors are quantized to a particular sub-pixel resolution. While this representation is chosen because it can be compressed efficiently, it does not capture complex and fine motion. In contrast, our algorithm has the full flexibility to distribute the bandwidth so that areas that matter more have arbitrarily sophisticated motion boundaries at an arbitrary flow precision, while unimportant areas are represented very efficiently. See comparisons in Figure~\ref{fig:flows_comparison}.

\vspace{-0.05in}
\paragraph{Multi-flow representation.} Consider a video of a train moving behind fine branches of a tree. Such a scene is highly inefficient to represent with traditional systems that use a single flow map, as there are small occlusion patterns that break the flow. Furthermore, the occluded content will have to be synthesized again once it reappears. We propose a representation that allows our method the flexibility to decompose a complex scene into a \emph{mixture of multiple simple flows} and preserve occluded content.

\vspace{-0.05in}
\paragraph{Spatial rate control.} It is critical for any video compression approach to feature a mechanism for assigning different bitrates at different spatial locations for each frame. In ML-based codec modeling, it has been challenging to construct a \emph{single model} which supports $R$ multiple bitrates, and achieves the same results as $R$ \emph{separate, individual models} each trained exclusively for one of the bitrates. In this work we present a framework for ML-driven spatial rate control which meets this requirement.

\subsection{Related Work}\label{sec:related_work}
\vspace{-0.05in}
\paragraph{ML-based image compression}
In the last two years, we have seen a great surge of ML-based image compression approaches \cite{gregor2011learning,toderici2015variable,toderici2016full,balle2016opt,balle2016end,NIPS2016_6542,pmlr-v48-larsen16,twitter2016,rippel17,johnston2017improved,NIPS2017_6714,li2017learning,balle2018variational,agustsson2018generative,dumas2018autoencoder,mentzer2018conditional,minnen2018joint}. These learned approaches have been reinventing many of the hard-coded techniques developed in traditional image coding: the coding scheme, transformations into and out of a learned codespace, quality assessment, and so on. 

\vspace{-0.05in}
\paragraph{ML-based video compression.} 
To our knowledge, the only pre-existing end-to-end ML-based video compression approach is \cite{wu2018vcii}. It first encodes key frames, and proceeds to hierarchically interpolate the frames between them. This approach performs on par with AVC/H.264. 

\vspace{-0.05in}
\paragraph{Enhancement of traditional coding using ML.} 
There have been several important contributions demonstrating the effectiveness of replacing or enhancing different components of traditional codecs with counterparts based on neural networks. These include improved motion compensation and interpolation \cite{yan2017convolutional,huo2018convolutional,zhao2018cnn,mathieu2015deep}, intra-prediction coding \cite{song2017neural}, post-processing refinement \cite{cavigelli2017cas,yang2017decoder,li2017cnn,yang2017enhancing,wang2017novel,yang2018multi,song2018practical,he2018enhancing}, and rate control \cite{li2017convolutional}.

\vspace{-0.05in}
\paragraph{Optical flow estimation.}\label{sec:flow_estimation}
The problem of \emph{optical flow estimation} has been widely studied over the years, with thousands of solutions developed using tools from partial differential equations \cite[\dots]{horn1981determining,lucas1981iterative,fleet2006optical,fortun2015optical} to, more recently, machine learning \cite[\dots]{weinzaepfel2013deepflow,fischer2015flownet,ilg2017flownet,ren2017unsupervised}. 

Given two similar frames $\rmbx_1, \rmbx_2\in\reals^{C\times H\times W}$, the task is to construct an optical flow field $\rmbf^* \in\reals^{2\times H\times W}$ of horizontal and vertical displacements, spatially ``shuffling'' values from $\rmbx_1$ to best match $\rmbx_2$. This can be more concretely written as
\begin{align}
\rmbf^* = \min_{\rmbf} \scL\left( \rmbx_2, \rmbF(\rmbx_1, \rmbf) \right) + \lambda\scR(\rmbf) \nonumber
\end{align}
for some metric $\scL(\cdot, \cdot)$, smoothness regularization $\scR(\cdot)$ and where $\rmbF(\cdot, \cdot)$ is the \emph{inverse optical flow} operator $\left[\rmbF(\rmbx, \rmbf)\right]_{chw} = \rmx_{c, h + \rmf_{1hw}, w + \rmf_{2hw}}$. Note that while $h, w$ are integer indices, $\rmf_{1hw}, \rmf_{2hw}$ can be real-valued, and so the right-hand side is computed using lattice interpolation. In this work we strictly discuss \emph{inverse} flow, but often simply write ``flow'' for brevity.

\subsection{Paper organization}\label{sec:paper_organization}
The paper is organized in the following way:
\begin{itemize}
\item In Section \ref{sec:architecture}, we motivate the overall design of our model, and present its architecture.\vspace{-0.05in}
\item In Section \ref{sec:coding_procedure}, we describe our coding procedure of generating variable-length bitstreams from the fixed-size codelayer tensors.\vspace{-0.05in}
\item In Section \ref{sec:spatial_rate_control}, we present our framework for ML-based spatial rate control, and discuss how we train/deploy it. \vspace{-0.15in}
\item \vspace{-0.05in} In Section \ref{sec:results}, we discuss our training/evaluation procedures, and present the results of our benchmarking and ablation studies. \vspace{-0.05in}
\end{itemize}


\section{Model Architecture}\label{sec:architecture}
\vspace{-0.05in}
\paragraph{Notation.} We seek to encode a video with frames $\rmbx_1, \ldots, \rmbx_T\in\reals^{3\times H\times W}$. Throughout this section, we discuss different strategies for video model construction. At a high level, all video coding models share the generic input-output structure in the below pseudo-code.
\vspace{-0.05in}
\begin{algorithm}[h]
\renewcommand{\thealgorithm}{}
\caption{Video coder structure for time step $t$}
\begin{algorithmic}[1]
  \REQUIRE 
  \STATE Target frame $\rmbx_t\in\reals^{3\times H\times W}$
  \STATE Previous state $\rmbS_{t-1}$
\end{algorithmic}

\begin{algorithmic}[1]
  \ENSURE 
  \STATE Bitstream $\rmbe_t\in \{0, 1\}^{\ell(\rmbe)}$ to be transmitted
  \STATE Frame reconstruction $\rmbhx_t\in\reals^{3\times H\times W}$
  \STATE Updated state $\rmbS_t$
\end{algorithmic}
\end{algorithm}
\vspace{-0.1in}

The state $\rmbS_t$ is made of one or more tensors, and intuitively corresponds to some \emph{prior memory} propagated from frame to frame. This concept will be clarified below.

\subsection{State propagator}\label{sec:state_propagator}
\vspace{-0.05in}
To motivate and provide intuition behind the final architecture, we present a sequence of steps illustrating how a traditional video encoding pipeline can be progressively adapted into an ML-based pipeline that is increasingly more general (and cleaner).

\vspace{-0.05in}
\paragraph{Step \#1: ML formulation of the flow-residual paradigm.}
Our initial approach is to simulate the traditional flow-residual pipeline featured by the existing codecs (see Section \ref{sec:existing_codecs}), using building blocks from our ML toolbox. We first construct a learnable \emph{flow estimator} network $\rmbM(\cdot)$, which outputs an (inverse) flow $\rmbf_t\in\reals^{2\times H\times W}$ motion-compensating the last reconstructed frame $\rmbhx_{t-1}$ towards the current target $\rmbx_t$. 

We then construct a learnable \emph{flow compressor} with encoder $\rmbE_f$ and decoder $\rmbD_f$ networks, which auto-encodes $\rmbf_t$ through a low-bandwidth bottleneck and reconstructs it as $\rmbhf_t$. Traditional codecs further increase the coding efficiency of the flow by encoding only the \emph{difference} $\rmbf_t - \rmbhf_{t-1}$ from the previously-reconstructed flow.

Next, we use our reconstruction of the flow to compute a motion-compensated reconstruction of the frame itself as $\rmbm_t = \rmbF(\rmbhx_{t-1}, \rmbhf_t)$, where $\rmbF(\cdot, \cdot)$ denotes the inverse optical flow operator (see Section \ref{sec:flow_estimation}). 

Finally, we build a \emph{residual compressor} with learnable encoder $\rmbE_r(\cdot)$ and decoder $\rmbD_r(\cdot)$ networks to auto-encode the residual $\bDelta_t = \rmbx_t - \rmbm_t$. Any state-of-the-art ML-based image compression architectures can be used for the core encoders/decoders of the flow and residual compressors. 

\begin{figure}[t]
  \centering
  \includegraphics[width=\columnwidth,trim=0cm 0cm 0cm 0cm,clip]{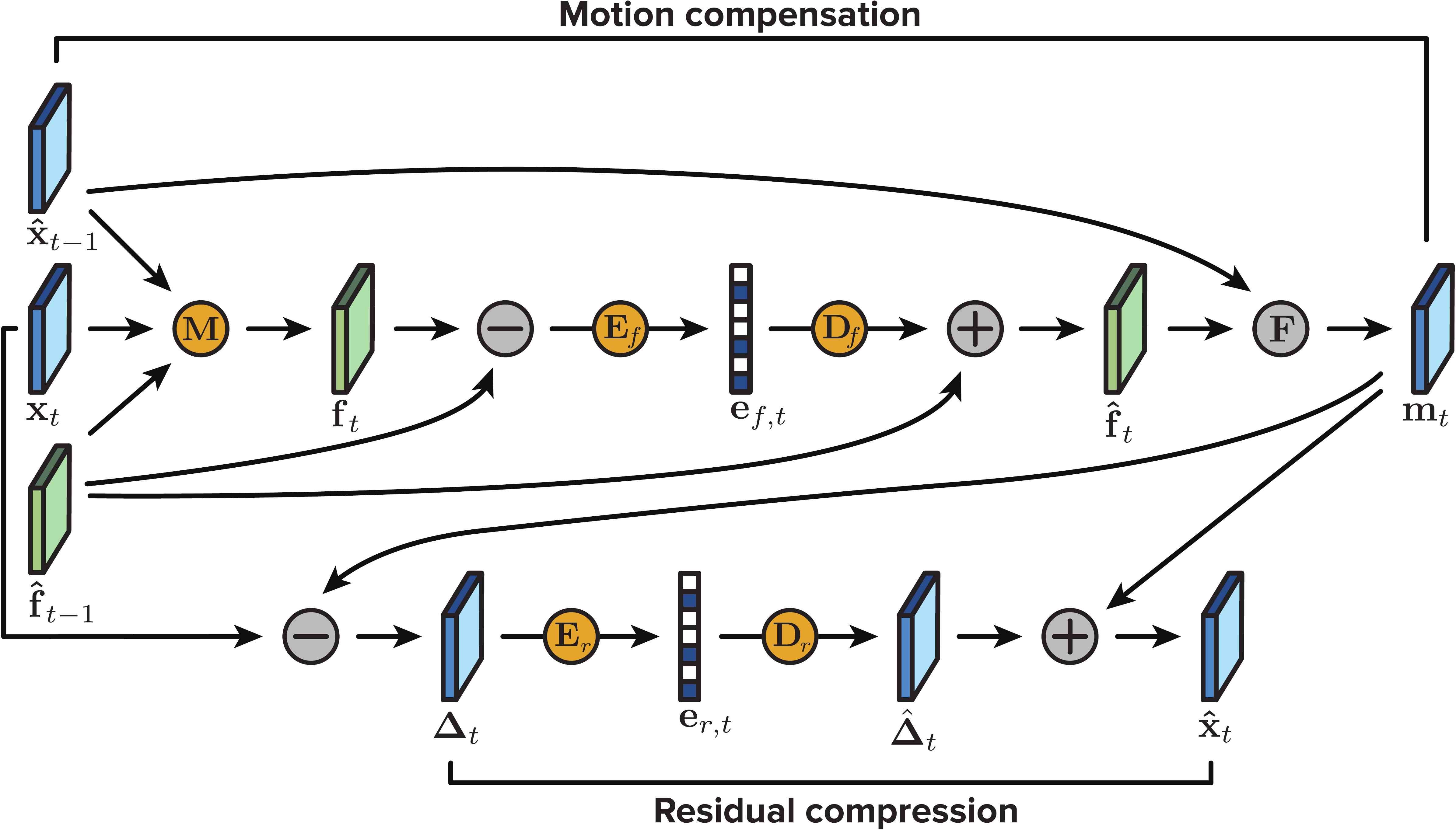}
  \caption{Graph of Step \#1, which formulates the traditional flow-residual pipeline using tools from ML. {\bf Blue} tensors correspond to frames and {\bf green} to flows, both embedded in raw pixel space. {\bf Yellow} operators are learnable networks, and {\bf gray} operators are hard-coded differentiable functions. $\rmbM(\cdot)$ is a flow estimator, and $\rmbF(\cdot, \cdot)$ is the optical flow operator described in Section \ref{sec:flow_estimation}.}
  \label{fig:step_1}
\vspace{-0.05in}
\end{figure}

See Figure~\ref{fig:step_1} for a visualization of this graph. While this setup generalizes the traditional approach via end-to-end learning, it still suffers from several important impediments, which we describe and alleviate in the next steps.

\begin{figure}[b]
  \centering
  \includegraphics[width=\columnwidth,trim=0cm 0cm 0cm 0cm,clip]{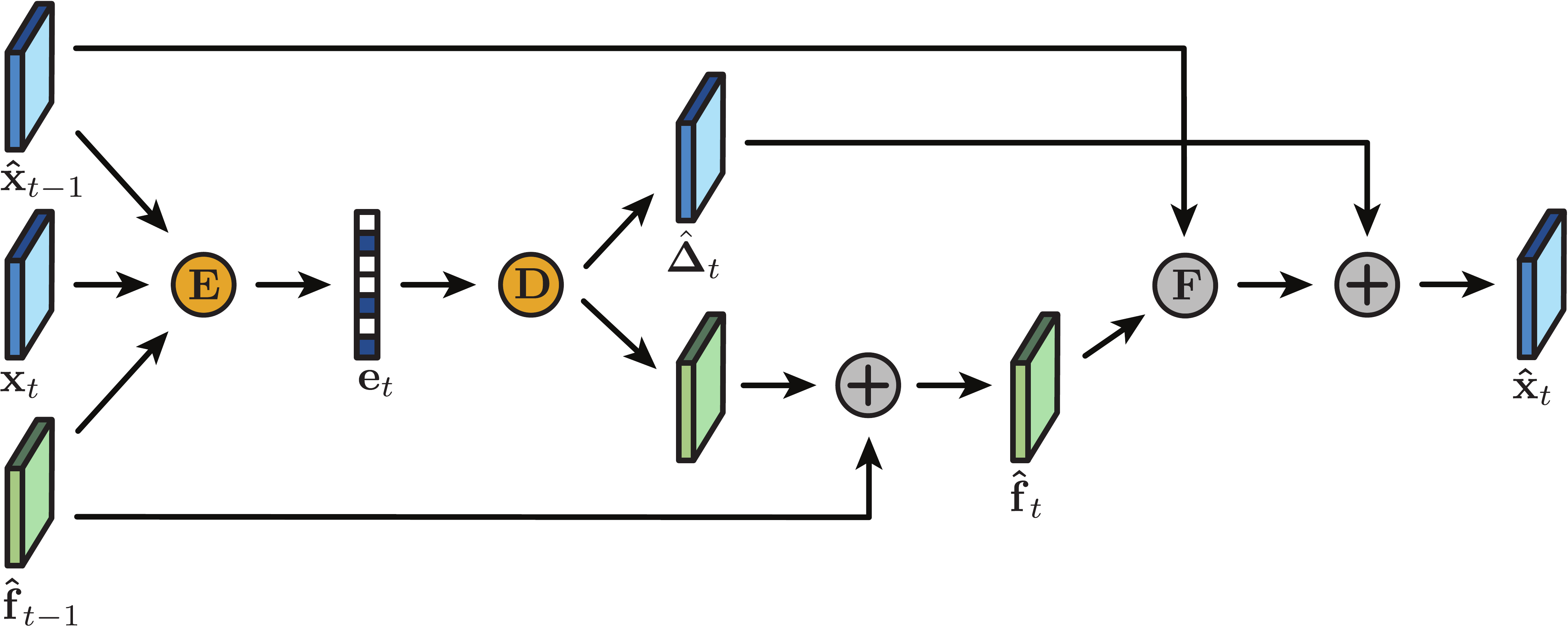}
  \caption{Graph of Step \#2, which generalizes the traditional architecture of Step \#1 by jointly compressing the flow and residual.}
  \label{fig:step_2}
\vspace{-0.05in}
\end{figure}

In the previous step, we encoded the flow and residual separately through distinct codes. Instead, it is advantageous in many ways to compress them jointly through a single bottleneck: this removes redundancies among them, and allows the model to automatically ascertain how to distribute bandwidth among them as function of input complexity. To that end, we consolidate to a single encoder $\rmbE(\cdot)$ network and single decoder $\rmbD(\cdot)$ network. See Figure~\ref{fig:step_2} for a graph of this architecture.

\vspace{-0.05in}
\paragraph{Step \#3: Propagation of a learned state.}
We now observe that all prior memory being propagated from frame to frame is represented strictly through the previously reconstructed frame $\rmbhx_{t-1}$ and flow $\rmbhf_{t-1}$, both embedded in raw pixel space. These representations are not only computationally inefficient, but also highly suboptimal in their expressiveness, as they can only characterize a very limited class of useful signals and cannot capture longer-term memory. Hence, it is greatly beneficial to define a generic and learnable \emph{state} $\rmbS_t$ of one or more tensors, and provide the model with a mechanism to automatically decide how to populate it and update it across time steps. 

\begin{figure}[t]
\vspace{-0.1in}
  \centering
  \includegraphics[width=\columnwidth,trim=0cm 0cm 0cm 0cm,clip]{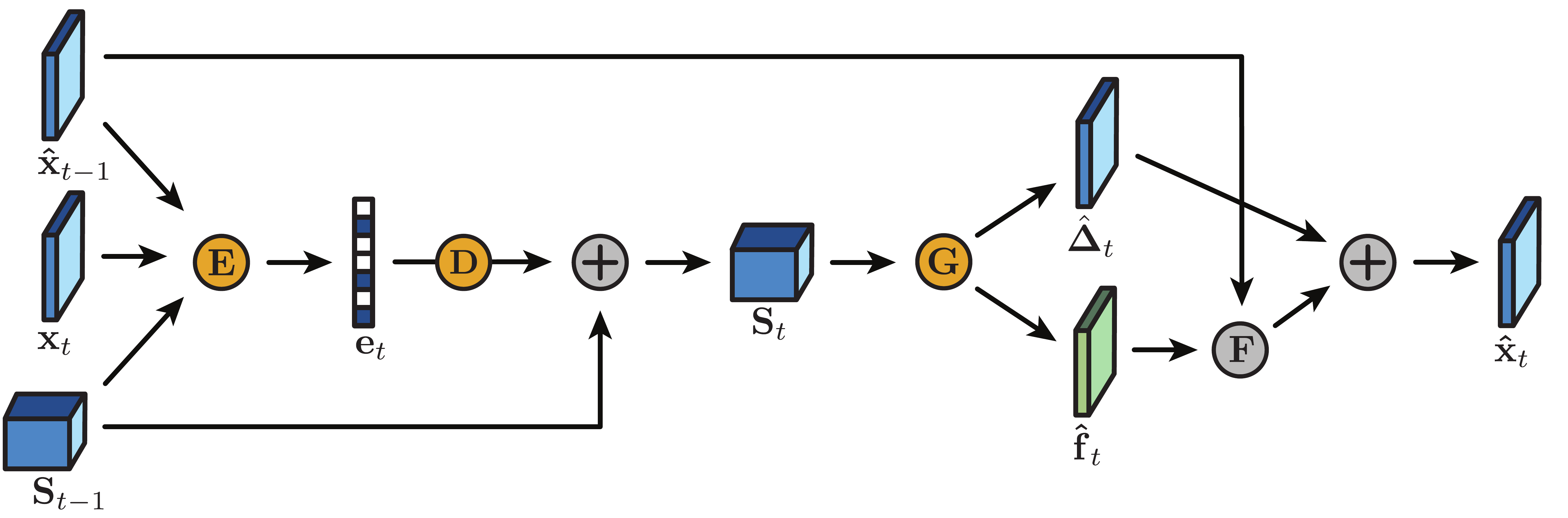}
  \caption{Graph of Step \#3. Rather than relying on reference frames and flows embedded in pixel space, we instead propagate a generalized state containing information learned by the model.}
  \label{fig:step_3}
\end{figure}

Our state propagation can be understood as an extension of a recursive neural network (RNN), where $\rmbS_t$ accumulates temporal information through recursive updates. Unlike a traditional RNN, updates to $\rmbS_t$ must pass through a low-bandwidth bottleneck, which we achieve through integration with modules for encoding, decoding, bitstream compression, compensation, and so on. Each frame reconstruction $\rmbhx_t$ is computed from the updated state $\rmbS_t$ using a module we refer to as \emph{state-to-frame}, and denote as $\rmbG(\cdot)$.

We provide an example skeleton of this architecture in Figure~\ref{fig:step_3}. In Figure~\ref{fig:ablation_studies}, it can be seen that introducing a learned state results in 20-40\% bitrate savings.

\vspace{-0.05in}
\paragraph{Step \#4: Arbitrary compensation.}
We can further generalize the architecture proposed in the previous step. We observe that the form of compensation in $\rmbG(\cdot)$ still simulates the traditional flow-based approach, and hence is limited to compensation of simple translations only. That is, flow-based compensation only allows us to ``move pixels around'', but does not allow us change their actual values. However, since we have a tool for end-to-end training, we can now learn any arbitrary compensation beyond motion. Hence, we can generalize $\rmbG(\cdot)$ to generate multiple flows instead of a single one, as well as multiple reference frames to which the flows can be applied respectively.


\begin{figure*}[t!]
\subfigure[Target frame $\rmbx_t$.]{
  \includegraphics[width=0.243\textwidth,trim=0cm 0cm 0cm 0cm,clip]{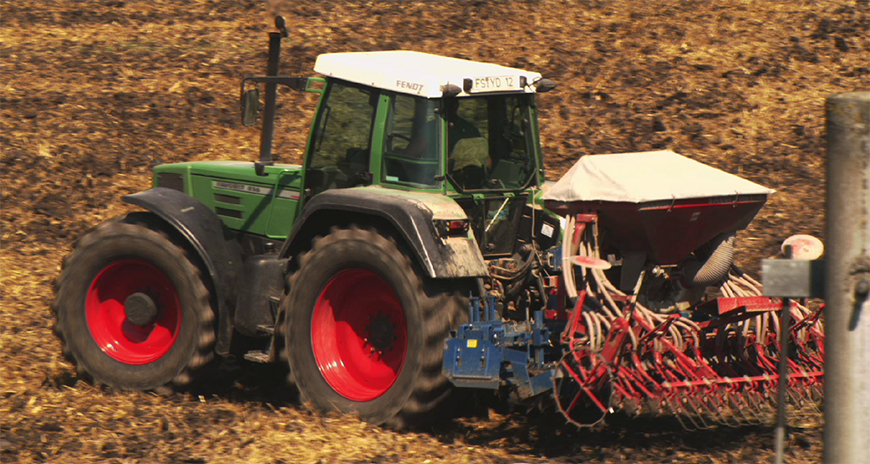}
}\hspace{-0.06in}
\subfigure[Target - previous $\rmbx_t - \rmbx_{t-1}$]{
  \includegraphics[width=0.243\textwidth,trim=0cm 0cm 0cm 0cm,clip]{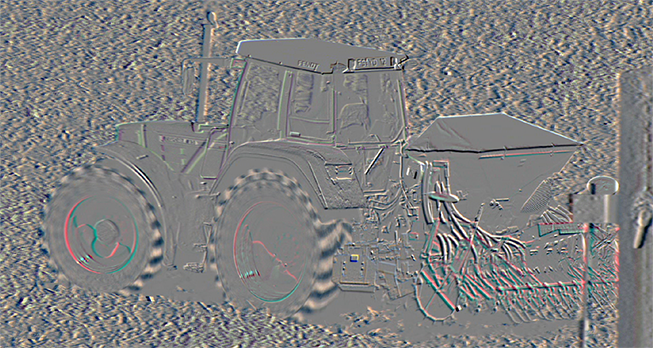}%
}\hspace{-0.03in}
\subfigure[Output optical flow $\rmbhf_t$.]{
  \includegraphics[width=0.243\textwidth,trim=0cm 0cm 0cm 0cm,clip]{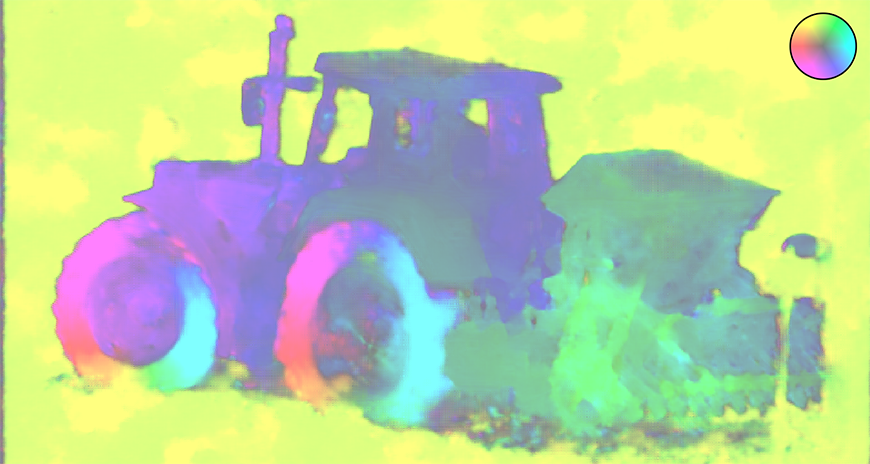}
}\hspace{-0.05in}
\subfigure[Residual $\rmbx_t - \rmbm_t$.]{
  \includegraphics[width=0.243\textwidth,trim=0cm 0cm 0cm 0cm,clip]{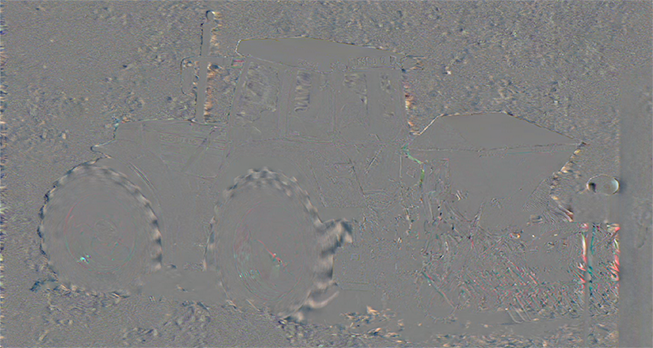}
}

\subfigure[Output reconstruction $\rmbhx_t$.]{
  \includegraphics[width=0.243\textwidth,trim=0cm 0cm 0cm 0cm,clip]{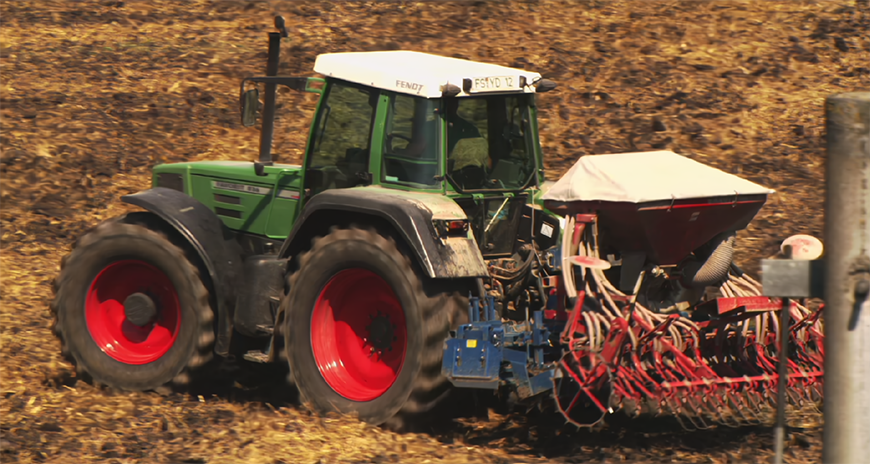}%
}\hspace{-0.03in}
\subfigure[Final error $\rmbx_t - \rmbhx_t$.]{
  \includegraphics[width=0.243\textwidth,trim=0cm 0cm 0cm 0cm,clip]{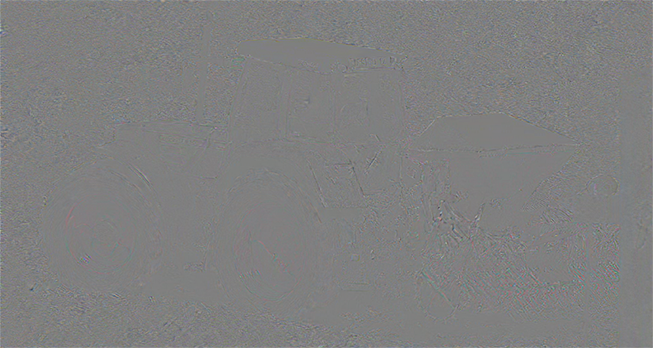}%
}\hspace{-0.03in}
\subfigure[MS-SSIM map.]{
  \includegraphics[width=0.243\textwidth,trim=0cm 0cm 0cm 0cm,clip]{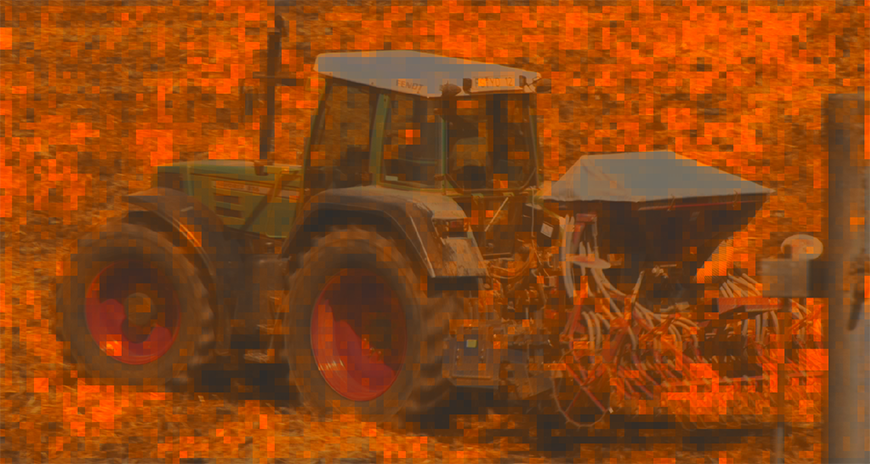}
}\hspace{-0.05in}
\subfigure[Output bitrate map.]{
  \includegraphics[width=0.243\textwidth,trim=0cm 0cm 0cm 0cm,clip]{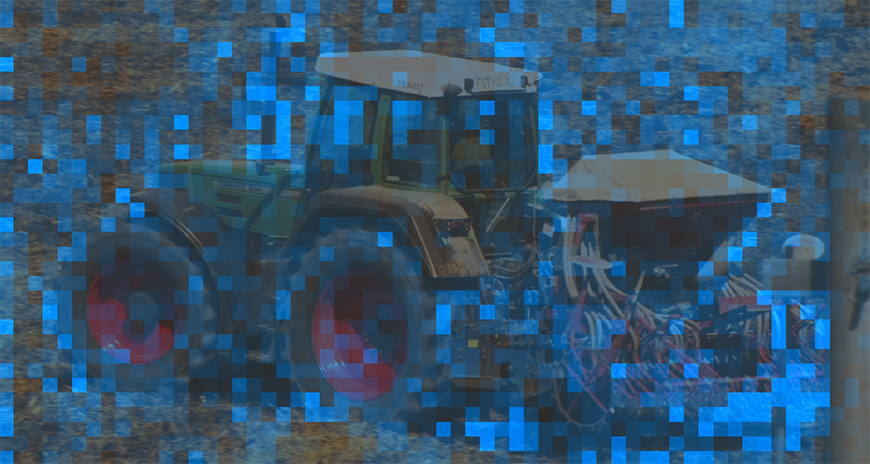}
}

\caption{Visualization of intermediate outputs for an example video from the Xiph HD dataset. {\bf (a)} The original target frame. {\bf (b)} The difference between the target and previous frame. There are several distinct types of motion: camera pan, turning wheels, and moving tractor. {\bf (c)} The output optical flow map as produced by the algorithm, compensating for the motion patterns as described in (b). {\bf (d)} The leftover residual following motion compensation. {\bf (e)} Output after addition of the residual reconstruction. {\bf (f)} The difference between the target and its final reconstruction. {\bf (g)} A map of the errors between the target and its reconstruction, as evaluated by MS-SSIM. Brighter color indicates larger error. {\bf (h)} A map of the bitrates assigned by the spatial rate controller as function of spatial location. Brighter color indicates higher bitrate.}
\label{fig:src_eyecandy}
\end{figure*}

\vspace{-0.05in}
\section{Coding procedure}\label{sec:coding_procedure}
\vspace{-0.05in}
We assume we have applied our encoder network, and have reached a fixed-size tensor $\rmbc\in[-1, 1]^{C\times Y\times X}$ (we omit timestamps for notational clarity). The goal of the coding procedure is to map $\rmbc$ to a bitstream $\rmbe\in\{0, 1\}^{\ell(\rmbe)}$ with variable length $\ell(\rmbe)$. The coder is expected to achieve high coding efficiency by exploiting redundancy injected into $\rmbc$ by the regularizer (see below) through the course of training. We follow the coding procedure described in detail in \cite{rippel17} and summarized in this section.

\vspace{-0.05in}
\paragraph{Bitplane decomposition.} 
We first transform $\rmbc$ into a binary tensor $\rmbb\in\{0, 1\}^{B\times C\times Y\times X}$ by decomposing it into $B$ bitplanes. This operation transformation maps each value $\rmc_{chw}$ into its binary expansion $\rmb_{1chw}, \ldots, \rmb_{Bchw}$ of $B$ bits. This is a lossy operation, since the precision of each value is truncated. In practice, we use $B=6$.

\vspace{-0.05in}
\paragraph{Adaptive entropy coding (AEC).}
The AEC maps the binary tensor $\rmbb$ into a bitstream $\rmbe$. We train a classifier to compute the probability of activation $\bbP[\rmb_{bcyx} = 1\,|\,C]$ for each bit value $\rmb_{bcyx}$ conditioned on some \emph{context} $C$. The context consists of values of neighboring pre-transmitted bits, leveraging structure within and across bitplanes.

\vspace{-0.05in}
\paragraph{Adaptive codelength regularization.}
The regularizer is designed to reduce the entropy content of $\rmbb$, in a way that can be leveraged by the entropy coder. In particular, it shapes the distribution of elements of the quantized codelayer $\rmbhc$ to feature an increasing degree of sparsity as function of bitplane index. This is done with the functional form
\begin{align}
\scR(\rmbhc) = \frac{\alpha_i}{CYX} \sum_{cyx} \log\left|\rmhc_{cyx}\right|\nonumber
\end{align}
for iteration $i$ and scalar $\alpha_i$. The choice of $\alpha_i$ allows training the mean codelength to match a target bitcount $\bbE_{\rmbhc}[\ell(\rmbe)] \longrightarrow \ell^{\textrm{target}}$. Specifically, during training, we use the coder to monitor the average codelength. We then modulate $\alpha_i$ using a feedback loop as function of the discrepancy between the target codelength and its observed value.


\section{Spatial Rate Control}\label{sec:spatial_rate_control}
It is critical for any video compression approach to include support for \emph{spatial rate control} --- namely, the ability to independently assign arbitrary bitrates $\rmr_{h, w}$ at different spatial locations across each frame. A \emph{rate controller} algorithm then determines appropriate values for these rates, as function of a variety of factors: spatiotemporal reconstruction complexity, network conditions, quality guarantees, and so on. 

\vspace{-0.05in}
\paragraph{Why not use a single-bitrate model?} Most of the ML-based image compression approaches train many individual models --- one for each point on the R-D curve \cite[\dots]{balle2016opt,rippel17,balle2018variational}. It is tempting to extend this formulation to video coding, and use a model which codes at a fixed bitrate level. However, one will quickly discover that this leads to fast accumulation of error, due to variability of coding complexity over both space and time. Namely, for a given frame, areas that are hard to reconstruct at a high quality using our fixed bitrate budget, are going to be even more difficult in the next frame, since their quality will only degrade further. In the example in Figure~\ref{fig:src_eyecandy}, it can be seen that different bitrates are assigned adaptively as function of reconstruction complexity. In Figure~\ref{fig:ablation_studies}, it can be seen that introducing a spatial rate controller results in 10-20\% better compression.

Traditional video codecs enable rate control via variation of the quantization parameters: these control the numerical precision of the code at each spatial location, and hence provide a tradeoff between bitrate and accuracy. 

In ML-based compression schemes, however, it has been quite challenging to design a high-performing mechanism for rate control. Concretely, it is difficult to construct a \emph{single model} which supports $R$ multiple bitrates, and achieves the same results as $R$ \emph{separate, individual models} each trained exclusively for one of the bitrates. 

In Section \ref{sec:multiplexer}, we present a framework for spatial rate control in a neural network setting, and in Section \ref{sec:rate_controller} discuss our controller algorithm for rate assignment.

\begin{figure}[b!]
  \includegraphics[width=\columnwidth,trim=0cm 0cm 0cm 0cm,clip]{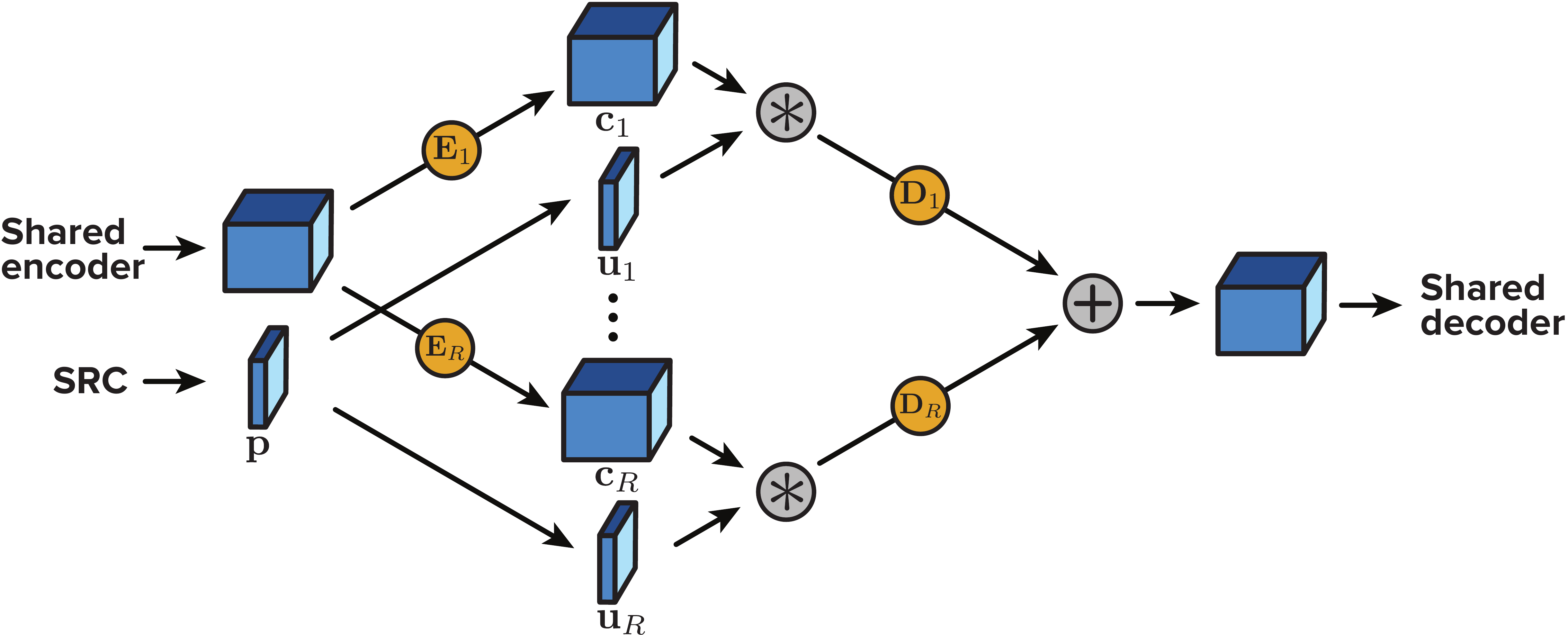}
  \caption{The architecture of the spatial multiplexer for rate control (Section \ref{sec:multiplexer}). At each location, a value is chosen from one of the $R$ codelayers, as function of the rate specified in the rate map $\rmbp$.}
  \label{fig:src_architecture}
\end{figure}

\begin{figure*}[t!]
\setlength{\columnsep}{0pt}
\centering

\begin{multicols}{2}
\centering
\qquad\quad\ \  {\bf CDVL SD}\\
\includegraphics[height=1.75in,trim=0cm 0cm 0cm 0cm,clip]{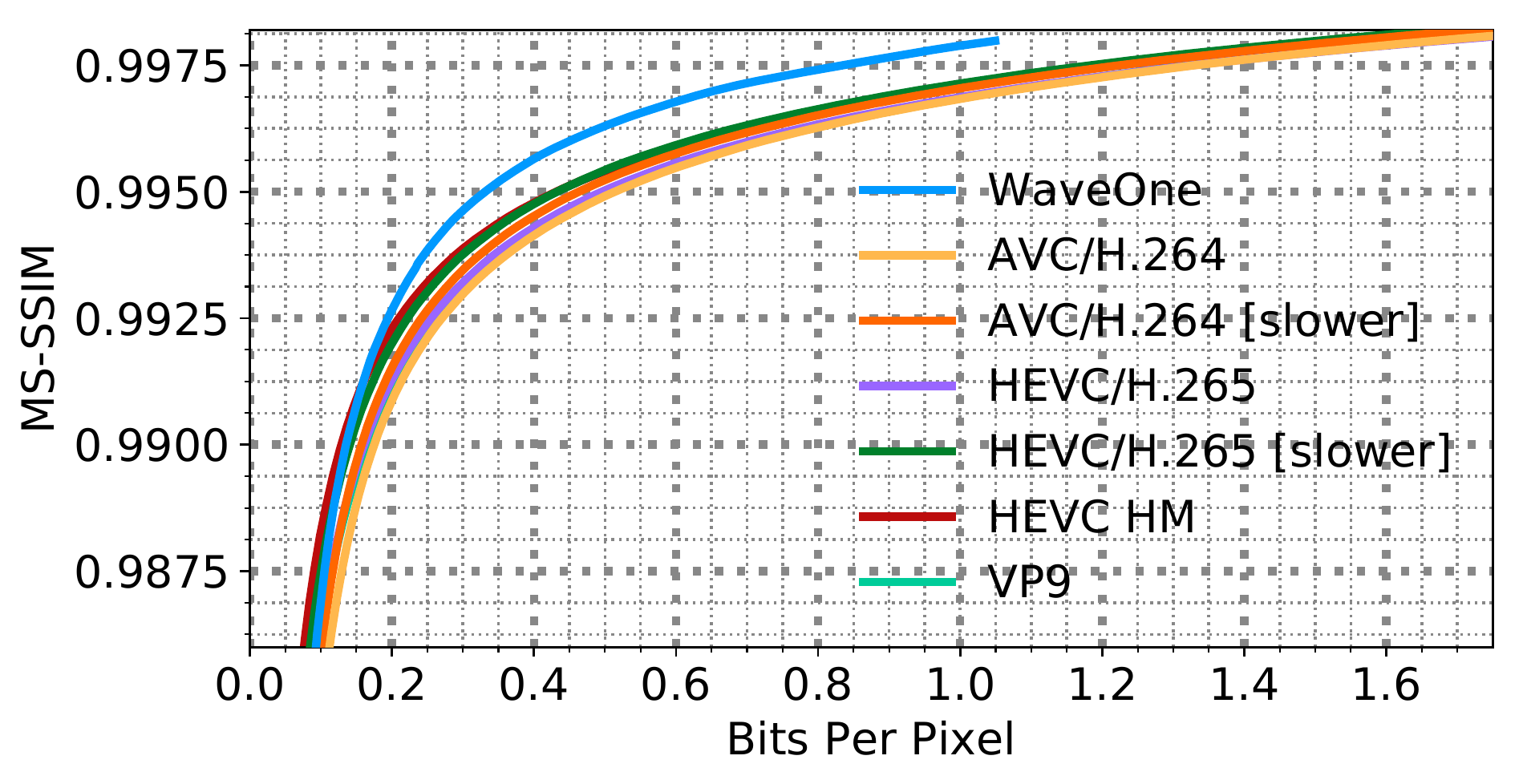}\\
\vspace{-0.12in}
\footnotesize
\qquad\quad\quad\ \ \begin{tabular}[t]{|r||r|r|r|r|r|}
    \hline
    {\bf MS-SSIM} & {\bf H.264} & {\bf H.265} & {\bf HEVC HM} & {\bf VP9} \\
    \hline\hline
    0.990  &    133\% &     122\% &   102\% &    123\% \\
    0.992  &    144\% &     133\% &   111\% &    133\% \\
    0.994  &    156\% &     148\% &   128\% &    152\% \\
    0.996  &    159\% &     152\% &   136\% &    154\% \\
    0.998  &    159\% &     160\% &   148\% &    160\% \\
    \hline
\end{tabular}
\normalsize

\qquad\quad{\bf Xiph HD}\\
\vspace{-0.03in}\includegraphics[height=1.75in,trim=0cm 0cm 0cm 0cm,clip]{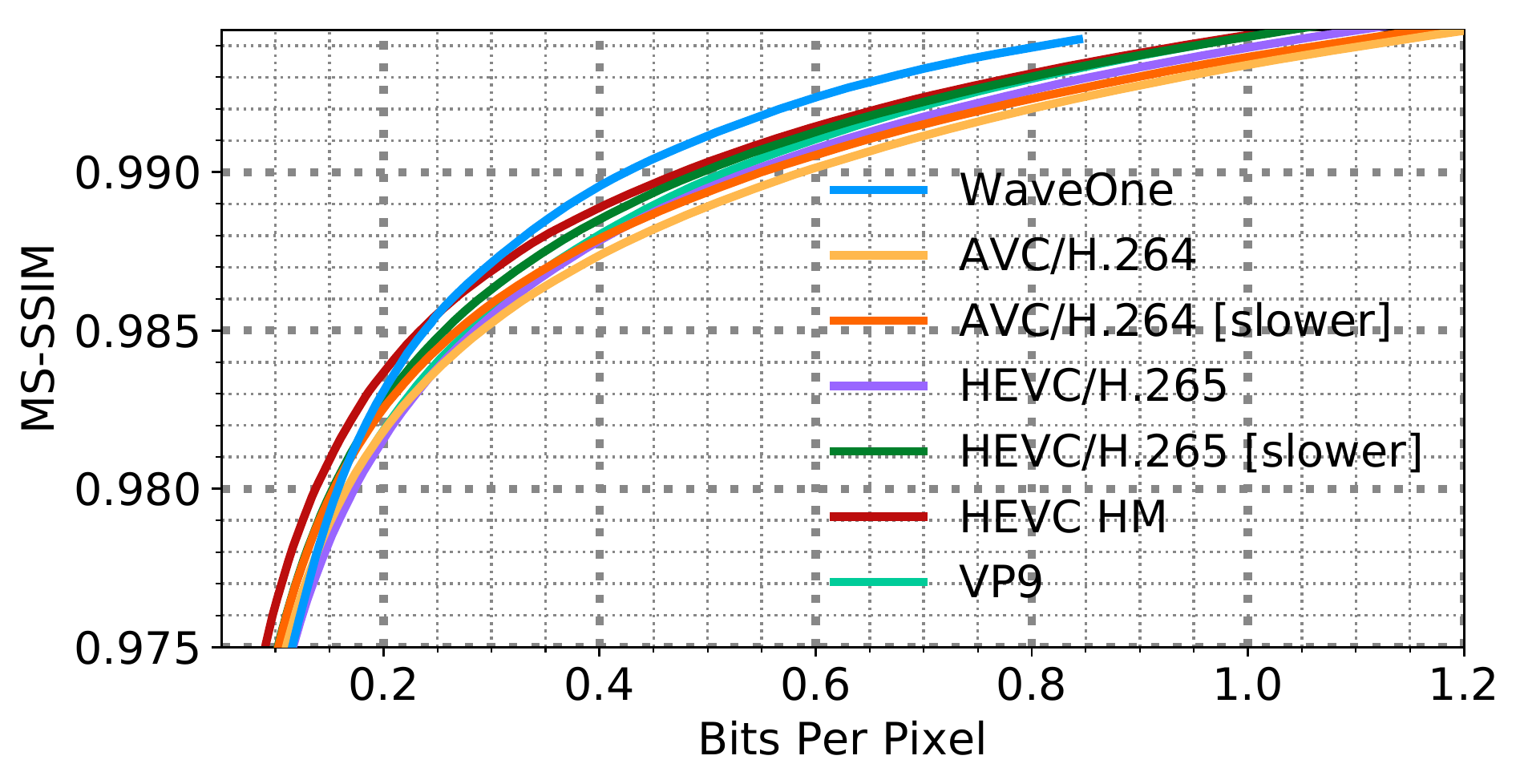} \\
\vspace{-0.12in}
\footnotesize
\qquad\quad\ \ \begin{tabular}[t]{|r||r|r|r|r|r|}
\hline
    {\bf MS-SSIM}  & {\bf H.264} & {\bf H.265} & {\bf HEVC HM} & {\bf VP9} \\
\hline\hline
            0.980  &    121\% &     112\% &   95\% &    102\% \\
            0.984  &    132\% &     120\% &  108\% &    112\% \\
            0.988  &    131\% &     116\% &  111\% &    111\% \\
            0.992  &    115\% &     108\% &  106\% &    105\% \\
            0.994  &    120\% &     113\% &  109\% &    108\% \\
\hline
\end{tabular}
\normalsize
\end{multicols}

\caption{Compression results for the CDVL SD and Xiph HD datasets. We benchmark against the default and slower presets of HEVC/H.265 and AVC/H.264, VP9, and the HEVC HM reference implementation, all in the low-latency setting (no B-frames). We tune each baseline codec to the best of our abilities. All details of the evaluation procedure can be found in Section \ref{sec:evaluation_procedure}. {\bf Top row:} Rate-distortion curves averaged across all videos for each dataset. {\bf Bottom row:} Average compressed sizes relative to WaveOne, for representative MS-SSIM levels covering the BPP range for each dataset.}
\setlength{\columnsep}{22.58437pt}
\label{fig:results}
\vspace{-0.1in}
\end{figure*}

\subsection{Spatial multiplexing framework}\label{sec:multiplexer}
\vspace{-0.05in}
Here we construct a mechanism which assigns variable bitrates across different spatial locations for each video frame. Concretely, we assume our input is a spatial map of integer rates $\rmbp\in\{1, 2, \ldots, R\}^{Y\times X}$. Our goal is to construct a model which can arbitrarily vary the BPP/quality at each location $(y, x)$, as function of the chosen rate $\rmp_{yx}$. 

To that end, we generalize our model featuring a single codelayer $\rmbc$ to instead support $R$ distinct codelayers $\rmbc_r\in\reals^{C_r\times Y_r\times X_r}$. Each codelayer is associated with a different rate, and is trained with a distinct entropy coder and codelength regularization (see Section \ref{sec:coding_procedure}) to match a different target codelength $\ell^{\textrm{target}}_r$. 

Our rate map $\rmbp$ then specifies which codelayer is active at each spatial location. In particular, we map $\rmbp$ into $R$ binary masks $\rmbu_r \in\{0, 1\}^{C_r\times Y_r\times X_r}$, one for each codelayer, of which a single one is active at each spatial location:
\begin{align}
\rmu_{r, cyx} = \bbI_{\rmp_{yx} = r}\,, \qquad r=1,\ldots,R\nonumber
\end{align}
where $\bbI$ is the indicator function. Each map $\rmbu_r$ masks codelayer $\rmbc_r$ during entropy coding. The final bitstream then corresponds to encodings of all the active values in each codelayer, as well as the rate mask itself (since it must be available on the decoder side as well).

In terms of the architecture, towards the end of the encoding pipeline, the encoder is split into $R$ branches $\rmbE_1, \ldots, \rmbE_R$, each mapping to a corresponding codelayer. Each decoder $\rmbD_r$ then performs the inverse operation, mapping each masked codelayer back to a common space in which they are summed (see Figure~\ref{fig:src_architecture}). To avoid incurring considerable computational overhead, we choose the individual codelayer branches to be very lightweight: each encoder/decoder branch consists of only a single convolution. 

In practice, we found that choosing target BPPs as $\ell^{\textrm{target}}_r = 0.01\times 1.5^r$ leads to a satisfactory distribution of bitrates. We train a total of 5 different models, each covering a different part of the BPP range. During training, we simply sample $\rmbp$ uniformly for each frame. Below we describe our use of the spatial multiplexer during deployment.

\vspace{-0.03in}
\subsection{Rate controller algorithm}\label{sec:rate_controller}
\vspace{-0.05in}
Video bitrate can be controlled in many ways, as function of the video's intended use. For example, it might be desirable to maintain a minimum guaranteed quality, or abide to a maximum bitrate to ensure low buffering under constraining network conditions (excellent overviews of rate control at \cite{wu2011rate,rate_control_site}). One common family of approaches is based on Lagrangian optimization, and revolves around assigning bitrates as function of an estimate of the \emph{slope} of the rate-distortion curve. This can be intuitively interpreted as maximizing the quality improvement per unit bit spent. 

Our rate controller is inspired by this idea. Concretely, during video encoding, we define some \emph{slope threshold} $\lambda$. For a given time step, for each spatial location $(y, x)$ and rate $r$, we estimate the slope $\frac{\scL_{r+1, yx} - \scL_{r, yx}} {\textrm{BPP}_{r+1, yx} - \textrm{BPP}_{r, yx}}$ of the local R-D curve, for some quality metric $\scL(\cdot, \cdot)$. We then choose our rate map $\rmbp$ such that at each spatial location, $\rmp_{yx}$ is the largest rate such that the slope is at least threshold $\lambda$.


\section{Results}\label{sec:results}
\subsection{Experimental setup}\label{sec:experimental_setup}

\vspace{-0.05in}
\paragraph{Training data.}
Our training set comprises high-definition action scenes downloaded from YouTube. We found these work well due to their relatively undistorted nature, and higher coding complexity. We train our model on $128\times 128$ video crops sampled uniformly spatiotemporally, filtering out clips which include scene cuts.

\vspace{-0.05in}
\paragraph{Training procedure.}
During training (and deployment) we encode the first frame using a learned image compressor; we found that the choice of this compressor does not matter much towards performance. We then unroll each video across 5 frames. We find diminishing returns from additional unrolling. We optimize the models with Adam \cite{kingma2014adam} with momentum of $0.9$ and learning rate of $2\times 10^{-4}$ which reduce by a factor of 5 twice during training. We use a batch size of $8$, and for a total of $400,000$ iterations.

\vspace{-0.05in}
\paragraph{Metrics and color space.}
For each encoded video, we measure BPP as the total file size, including all header information, averaged across all pixels in the video.

We penalize discrepancies between the final frame reconstructions $\rmbhx_t$ and their targets $\rmbx_t$ using the Multi-Scale Structural Similarity Index (MS-SSIM) \cite{wang2003multiscale}, which has been designed for and is known to match the human visual system significantly better than alternatives such as PSNR or $\ell_p$-type losses. We penalize distortions in all intermediate motion-compensated reconstructions using the Charbonnier loss, known to work well for flow-based distortions \cite{sun2010secrets}. 

Since the human visual system is considerably more sensitive to distortions in brightness than color, most existing videos codecs have been designed to operate in the YCbCr color space, and dedicate higher bandwidth to luminance over chrominance. Similarly, we represent all colors in the YCbCr domain, and weigh all metrics with Y, Cb, Cr component weights $6/8, 1/8, 1/8$.

\subsection{Evaluation procedure}\label{sec:evaluation_procedure}

\paragraph{Baseline codecs.}
We benchmark against all mainstream commercial codecs: HEVC/H.265, AVC/H.264, VP9, and the HEVC HM 16.0 reference implementation. We evaluate H.264 and H.265 in both the default preset of \code{medium}, as well as \code{slower}. We use FFmpeg for all codecs, apart from HM for which we use its official implementation. We tune all codecs to the best of our ability. To remove B-frames, we use H.264/5 with the \code{bframes=0} option, VP9 with \code{-auto-alt-ref 0 -lag-in-frames 0}, and use the HM \code{encoder\_lowdelay\_P\_main.cfg} profile. To maximize the performance of the baselines over the MS-SSIM metric, we tune them using the \code{-ssim} flag.

\paragraph{Video test sets.}
We benchmark all the above codecs on standard video test sets in SD and HD, frequently used for evaluation of video coding algorithms. In SD, we evaluate on a VGA resolution dataset from the Consumer Digital Video Library (CDVL)\footref{cdvl_sd}. This dataset has 34 videos with a total of 15,650 frames. In HD, we use the Xiph 1080p video dataset\footref{xiph}, with 22 videos and 11,680 frames. We center-crop all 1080p videos to height 1024 (for now, our approach requires each dimension to be divisible by 32). Lists of the videos in each dataset can be found in Appendices \ref{sec:appendix_cdvl}, \ref{sec:appendix_hd}.

\paragraph{Curve generation.}
Each video features a separate R-D curve computed from all available compression rates for a given codec: as a number of papers \cite{balle2016opt,rippel17} discuss in detail, different ways of summarizing these R-D curves can lead to very different results. In our evaluations, to compute a given curve, we sweep across values of the independent variable (such as bitrate). We interpolate the R-D curve for each video at this independent variable value, and average all the results across the dependent variable. To ensure accurate interpolation, we generate results for all available rates for each codec.

\subsection{Performance}
We present several different types of results:

\vspace{-0.05in}
\paragraph{Rate-distortion curves.} On the top row of Figure~\ref{fig:results}, we present the average MS-SSIM across all videos for each dataset and for each codec (Section \ref{sec:evaluation_procedure}), as function of BPP.

\vspace{-0.05in}
\paragraph{Relative compressed sizes.} On the bottom row of Figure~\ref{fig:results}, we present average file sizes relative to our approach for representative MS-SSIM values. For each MS-SSIM point, we average the BPP for all videos in the dataset and compute the ratio to our BPP. Note that for this comparison, we are constrained to use MS-SSIM values that are valid for all videos in the dataset, which is 0.990-0.998 for the SD dataset and 0.980-0.994 for the HD dataset.

\begin{figure}[t]
\centering
\includegraphics[width=\columnwidth,trim=0cm 0cm 0cm 0cm,clip]{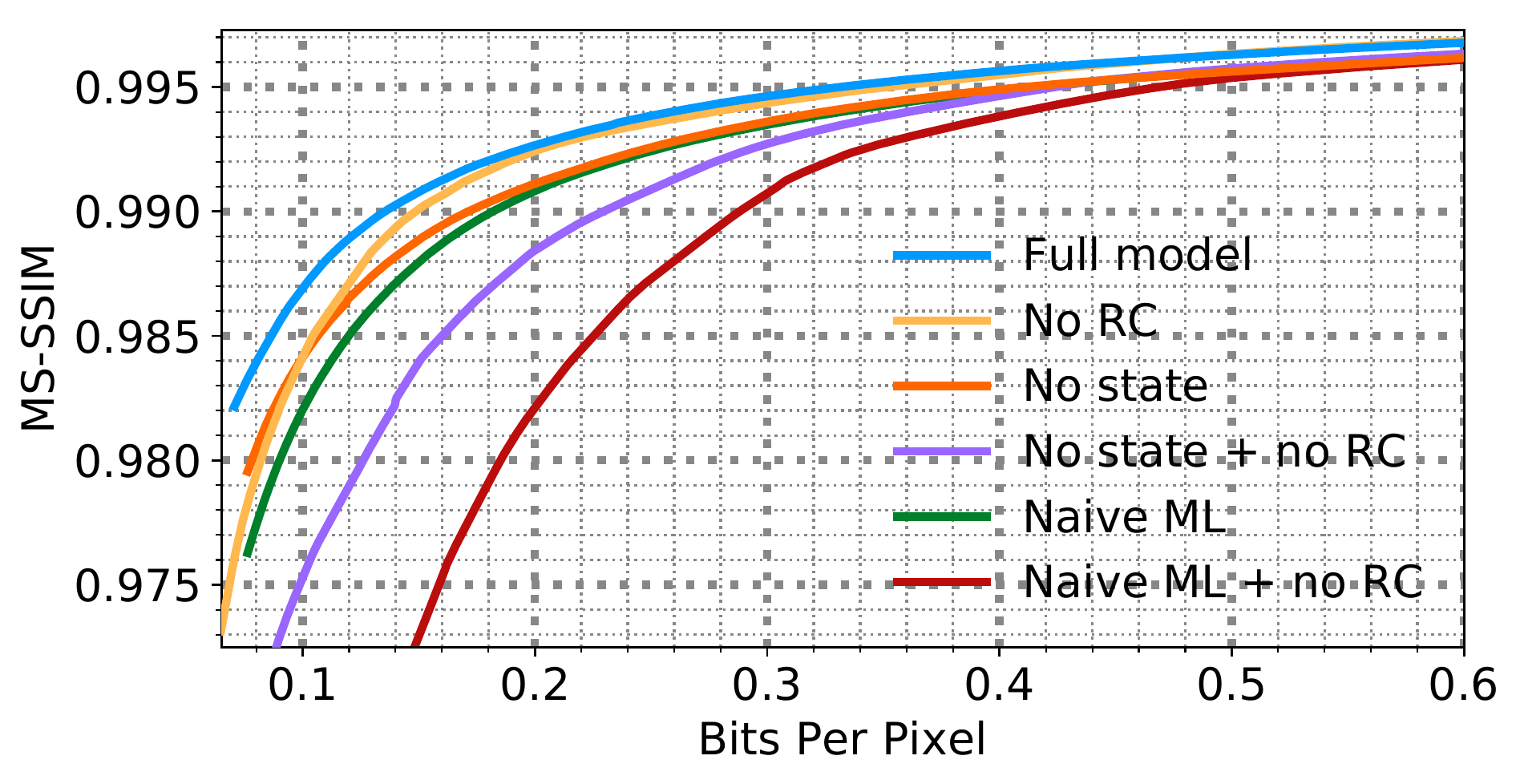}

\caption{Ablation studies demonstrating the impact of individual architectural components on performance on the CDVL SD dataset. Factors of variation include introduction of a learned state, use of flow-based motion compensation, and spatial rate control (all described in Sections \ref{sec:architecture} and \ref{sec:spatial_rate_control}).}
\label{fig:ablation_studies}
\end{figure}

\vspace{-0.05in}
\paragraph{Ablation studies.} In Figure~\ref{fig:ablation_studies}, we present relative performance of different models with and without different architectural components. The different configurations evaluated include:
\begin{itemize}
\item The full model presented in the paper,
\item The model described in Step \#2, using previous frames and flows as prior knowledge, but without learning an arbitrary state, and
\item A Na\"{i}ve ML model, which does not include a learned state, and reconstructs the target frame directly without any motion compensation.
\end{itemize}
We evaluate all the above models with and without the spatial rate control framework described in Section \ref{sec:spatial_rate_control}.

\stepcounter{footnote}
\footnotetext{The Consumer Digital Video Library can be found at \url{http://www.cdvl.org/}. To retrieve the SD videos, we searched for VGA resolution at original and excellent quality levels. There were a few instances of near-duplicate videos: in those cases we only retrieved the first.\label{cdvl_sd}}

\stepcounter{footnote}
\footnotetext{The Xiph test videos can be found at \url{https://media.xiph.org/video/derf/}. We used all videos with 1080p resolution.\label{xiph}}

\vspace{-0.05in}
\paragraph{Runtime.}
On an NVIDIA Tesla V100 and on VGA-sized videos, our decoder runs on average at a speed of around 10 frames/second, and encoder at around 2 frames/second irrespective of bitrate. However, our algorithm should be regarded as a \emph{reference implementation}: the current speed is not sufficient for real-time deployment, but is to be substantially improved in future work. For reference, on the same videos, the HEVC HM algorithm encodes at around 0.3 frames/second for low BPPs and 0.04 frames/second for high BPPs.

\section{Conclusion}
In this work we introduced the first ML-based video codec that outperforms all commercial codecs, across nearly the entire bitrate range in the low-latency mode. However, our presented approach only supports the low-latency mode. Two clear directions of future work are to increase the computational efficiency of the model to enable real-time coding, as well as extend the model to support temporal interpolation modes (i.e, using B-frames).

\paragraph{Acknowledgements.} We are grateful to Trevor Darrell, Sven Strohband, Michael Gelbart, Albert Azout, Bruno Olshausen and Vinod Khosla for meaningful discussions and input along the way. 

{\small
\bibliographystyle{ieee}
\bibliography{main}
}

\clearpage
\newpage
\begin{appendices}
\begin{multicols}{1} 
\title{Learned Video Compression: Supplementary Material}
\author{}
\maketitle
\end{multicols}


\section{Detailed description of test sets}
\subsection{CDVL SD}\label{sec:appendix_cdvl}
The Consumer Digital Video Library can be found at \url{http://www.cdvl.org/}. To retrieve the SD videos, we searched for VGA resolution at original and excellent quality levels. There were a few instances of near-duplicate videos: in those cases we only retrieved the first. All videos are listed below.

\begin{lstlisting}[tabsize=1,basicstyle=\scriptsize\ttfamily]
Bennet-Watt_BeeClose_VGA60fps
Bennet-Watt_BeeZoom_VGA60fps
Bennet-Watt_CattleDogs_VGA60fps
Bennet-Watt_DecantWine_VGA60fps
Bennet-Watt_ElephantZoom_VGA60fps
Bennet-Watt_FlockSunset_VGA60fps
ntia_bpit1-vga_original
ntia_bpit2-vga_original
ntia_bpit3-vga_original
ntia_bpit4-vga_original
ntia_bpit5-vga_original
ntia_cardark-vga_original
ntia_cargas-vga_original
ntia_catjoke-vga_original
ntia_cchart1-vga_original
ntia_diner-vga_original
ntia_drmfeet-vga_original
ntia_drmside-vga_original
ntia_fish1-vga_original
ntia_fish5-vga_original
NTIA_FlamencoDancers_VGA60fps
NTIA_FlamencoShoes_VGA60fps
ntia_flower1-vga_original
ntia_overview1-vga_original
ntia_rfdev1-vga_original
ntia_schart1-vga_original
ntia_spectrum1-vga_original
ntia_store1-vga_original
ntia_street1-vga_original
NTIA_TheFootDrummer_VGA60fps
NTIA_TheFootPan_VGA60fps
NTIA_TheFootPiano_VGA60fps
NTIA_WaveRocks_VGA60fps
ntia_wboard1-vga_original
\end{lstlisting}

\subsection{Xiph HD}\label{sec:appendix_hd}
The Xiph test videos can be found at \url{https://media.xiph.org/video/derf/}. We used all videos with 1080p resolution.

\begin{lstlisting}[tabsize=1,basicstyle=\scriptsize\ttfamily]
aspen_1080p.y4m
blue_sky_1080p25.y4m
controlled_burn_1080p.y4m
crowd_run_1080p50.y4m
dinner_1080p30.y4m
ducks_take_off_1080p50.y4m
in_to_tree_1080p50.y4m
life_1080p30.y4m
old_town_cross_1080p50.y4m
park_joy_1080p50.y4m
pedestrian_area_1080p25.y4m
red_kayak_1080p.y4m
riverbed_1080p25.y4m
rush_field_cuts_1080p.y4m
rush_hour_1080p25.y4m
snow_mnt_1080p.y4m
speed_bag_1080p.y4m
station2_1080p25.y4m
sunflower_1080p25.y4m
touchdown_pass_1080p.y4m
tractor_1080p25.y4m
west_wind_easy_1080p.y4m
\end{lstlisting}
\end{appendices}

\end{document}